\documentclass[twocolumn]{svjour3}
\usepackage[latin1]{inputenc}
\usepackage[colorlinks=true,ps2pdf=true]{hyperref}
\usepackage[dvips]{graphics}
\usepackage{psfrag}
\usepackage{amsmath}
\usepackage{amssymb}
\usepackage[english]{babel}
\usepackage{xspace}
\usepackage{url}
\usepackage{theorem}
\newcommand{\etal}{\textit{et al.}~}	
\newcommand{\ie}{{i.e.},\ }		
\newcommand{\eg}{{e.g.},\ }		
\newcommand{\cf}{{cf.}\ }               
\def\rcs$#1${#1}

\newcommand{\term}[1]{\emph{#1}}		
\newcommand{\Z}{\mathbb{Z}}		
\newcommand{\sendright}[1]{$\xrightarrow{\quad #1\quad}$}
\newcommand{\sendleft}[1]{$\xleftarrow{\quad #1\quad}$}
\newcommand{\broadcastright}[1]{$\@Rightarrow{\quad #1\quad}$}
\newcommand{\broadcastleft}[1]{$\@Leftarrow{\quad #1\quad}$}

\newcommand{\xor}{\oplus}		
\newcommand{\assign}{\mathrel{:=}}	

\newcommand{\id}[1]{\ensuremath{\mathit{#1}}}	

\newcommand{\actor}{d}
\newcommand{\domain}{D}
\newcommand{\domains}{\mathcal{D}}
\newcommand{\curdomain}{\Gamma}
\newcommand{\caller}{\ensuremath{\mathit{caller}}}
\newcommand{\rfidtag}{t}
\newcommand{\rfidtagowner}{T}
\newcommand{\rfidtagowners}{\mathcal{T}}
\newcommand{\obj}{o}
\newcommand{\objmngrobj}{\Omega}
\newcommand{\class}{c}
\newcommand{\classowner}{C}

\newcommand{\method}{f}

\newcommand{\func}[1]{\textsc{\textbf{#1}}}
\newcommand{\objname}[1]{\textit{#1}}
\newcommand{\datetime}{\delta}
\newcommand{\maxtime}{\Delta}

\newcommand{\secparam}{\gamma}

\newcommand{\symkey}{k}
\newcommand{\pubkey}{PK}
\newcommand{\key}{sk}
\newcommand{\kta}{\symkey_a} 
\newcommand{\kMA}{\symkey_A} 
\newcommand{\kc}{\symkey_\class}
\newcommand{\KD}{\pubkey_\domain}
\newcommand{\kD}{{\key_\domain}}
\newcommand{\eka}{E} 
\newcommand{\epoch}{e}
\newcommand{\curepoch}{\epsilon}
\newcommand{\symkD}{\symkey_\domain}
\newcommand{\objmngrkey}{\symkey_\objmngrobj}

\newcommand{\pcfod}{\symkey_{\class,\method,\domain,\maxtime}}
\newcommand{\encid}{\id{id}} 

\newcommand{\enc}[2]{\{#2\}_{#1}}
\newcommand{\mac}[2]{[#2]_{#1}}
\newcommand{\macenc}[2]{[\{#2\}]_{#1}}
\newcommand{\now}{\mathit{now}}
\newcommand{\owner}{\mathit{owner}}
\newcommand{\accessset}{A}
\newcommand{\concat}{;}

\newcommand{\fixlistspacing}{}
\renewcommand{\descriptionlabel}[1]{\emph{#1}\hfil}

\journalname{None}

\def\makeheadbox{{%
\hbox to0pt{\vbox{\baselineskip=10dd\hrule\hbox
to\hsize{\vrule\kern3pt\vbox{\kern3pt
\hbox{\bfseries Preprint}
\hbox{\rcs$Id: sensorpets.tex 52 2010-02-25 14:02:07Z jhh $}
\kern3pt}\hfil\kern3pt\vrule}\hrule}%
\hss}}}

\date{\today}

\title{Practical Schemes For\\ Privacy \& Security Enhanced RFID}

\author{Jaap-Henk Hoepman \and Rieks Joosten}
\institute{Jaap-Henk Hoepman \at
TNO Information and Communication Technology,
  \email{jaap-henk.hoepman@tno.nl},
  and
  Institute for Computing and Information Sciences,
  Radboud University Nijmegen,
  \email{jhh@cs.ru.nl}
\and
Rieks Joosten \at
TNO Information and Communication Technology
  \email{rieks.joosten@tno.nl}
}

\begin{document}

\maketitle{}

\bibliographystyle{acm}

%

\hyphenation{a-na-ly-sis}
\hyphenation{meta-phor}

\begin{abstract}
Proper privacy protection in RFID systems is important. However, many of the
schemes known are impractical. Some use hash functions instead
of the more hardware efficient symmetric encryption schemes as a 
cryptographic primitive. Others incur a rather large
time penalty at the reader side, because the reader has to perform a 
key search over large tag key space. Moreover, they do not allow for dynamic,
fine-grained access control to the tag that cater for more complex usage
scenarios.

In this paper we investigate such scenarios, and propose a model and
corresponding privacy friendly protocols for efficient and fine-grained
management of access permissions to tags. In particular we propose an
efficient mutual authentication protocol between a tag and a reader that
achieves a reasonable level of privacy, using only symmetric key cryptography
on the tag, while not requiring a costly key-search algorithm at the reader
side. Moreover, our protocol is able to recover from stolen readers.
\end{abstract}

\section{Introduction}

Radio Frequency Identification (RFID) is a technology that allows to wirelessly
identify and collect data about a particular physical object from a
relatively short distance (depending on the technology used ranging from a few
centimeters up to several meters). The data is stored on so-called tags attached to
the object, and is collected using so-called readers. RFID tags can be
very small, can be attached invisibly to almost anything, and can transmit
potentially unique identifying information. Therefore,
proper privacy protection within
RFID based systems is of paramount
importance~\cite{garfinkel2005rfid-privacy,juels2006rfid-secpriv-survey}.

Yet RFID is also an enabler for the vision of an
Inter\-net-of-Things where the physical and the virtual become interconnected in
one single network. This will spark all kinds of applications
beyond our current imagination.
Some of these applications may be useful and
beneficial for individuals and society, others may be potentially very damaging
(to our personal liberties, or otherwise). 
It would be a waste, however, to abort such future innovations by mandating the
use of a kill-switch on all RFID tags\footnote{%
	As recommended in EC Recommendation (SEC(2009) 585/586) of 12.5.2009 on
	the implementation of privacy and 
	data protection principles in applications supported by radio-frequency
	identification.
} 
that will silence such a tag forever once it leaves the shop. Such a
kill-switch is a very coarse, all-or-nothing approach to protecting privacy. 
It would be far better to develop an approach that allows the user to have fine
grained and dynamic control over who can access his tags, and when.
The research reported on in this paper takes a step into that direction.

\subsection{State of the art}
\label{ssec-state}

Because of the privacy risk associated with the large scale use of RFID tags, 
many proposals exist to provide a certain level of privacy
protection for a particular application of RFID. We give a brief overview of
the state of the art, focusing on authentication and access control.

Early proposals use
relabelling of tag identifiers~\cite{sarma2002rfid-report},
or re-encryption 
techniques~\cite{juels2003squealing,avoine2004rfidbanknotes,golle2004reencryption}
that randomly encrypt the identifier from time to time, so that it can only be recovered by authorised readers, while being untraceable for others. 

Another approach is to implement some form of authentication between tag and
reader, and to allow only authorised tags to retrieve the tag identifier.
In a public key setting this would be easy, but RFID tags are generally
considered to be too resource poor to accommodate for that. 
Therefore, several identification and
authentication protocols using hash functions or symmetric key
cryptography have been
proposed~\cite{weis2003security,engberg2004zeroknowledge-rfid}. 
In particular,
Ohkubo, Suzuki, and Kinoshita~\cite{ohkubo2004hash-chains-rfid} present a
technique for achieving forward privacy in tags. This
property means that if an attacker compromises a tag,
i.e., learns its current state and its key, she is nonetheless
unable to identify the previous outputs of the same tag.
In their protocol, a tag has a unique identifier $\id{id}_i$, that is changed
every time the tag is queried by a reader. In fact, when queried for the $i$-th
time, the tag responds with $g(\id{id}_i)$ to the reader, and sets 
$\id{id}_{i+1} = h(\id{id}_i)$ immediately after that. 
Dimitirou~\cite{dimitriou2005lightweight-rfid-protocol} presents a similar
protocol, but that authenticates the tag as well.
In both cases, if all readers are \emph{on line}, connected with one central
database, the readers can be synchronised and the response of a tag can be
looked up immediately in the database. 
(Note that the database can keep a
shadow copy of $\id{id}_i$ and hence can precompute the next expected value
$g(h(\id{id}_i))$.) 
If not, or if synchronisation errors
occur, a search over all possible (initial) identifiers
(expanding hash chains)
is necessary.

In a symmetric key setting the reader cannot know the identifier of the tag
a priori, or obtain the identifier of the tag at the start of the protocol 
because of privacy concerns. One can give all readers and tags the
same symmetric key, but this has the obvious drawback that once the key of one
tag is stolen, the whole system is corrupted. To increase security, 
tags can be given separate keys, but then the reader must search the
right key to use for a particular tag.
This issue is not properly addressed
in Engberg's paper~\cite{engberg2004zeroknowledge-rfid}. It is unclear whether
in that paper tags share a single a key with a group of other tags, or that
each tag has a unique and private access key it only shares with the reader.
The core challenge is therefore to provide, possibly
efficient, trade offs and solutions for key search and key management. 
Molnar and Wagner~\cite{DBLP:conf/ccs/MolnarW04} (see also~\cite{DBLP:conf/percom/Dimitriou06})
propose to arrange keys in a tree
structure, where individual tags are associated with leaves in the tree, and
where each tag contains the keys on the path from its leaf to the root.
In subsequent work Molnar, Soppera, and Wagner~\cite{DBLP:conf/sacrypt/MolnarSW05}
explore ways in
which the sub-trees in their scheme may be associated with
individual tags. They also introduce the concept of delegation that allows a
tag owner to enable another party to access a tag over some period of time,
like for instance a fixed number of read operations.
In another approach, 
Avoine, Dysli, and Oechslin~\cite{DBLP:conf/sacrypt/AvoineDO05,DBLP:conf/percom/AvoineO05}
show how, similar to the the study of Hellman
to breaking symmetric keys, 
a time-memory trade off can be exploited to make the search for the key to use
more efficient.
We note that none of these systems are practical for RFID systems where millions of
tags possess unique secret keys.

Spiekermann~\etal\cite{spiekermann2009critical-rfid-pet}
observe that although there are many protocols and proposals for 
limiting access to RFID tags (either by killing them completely or by requiring
the reader to authenticate), few systems have been proposed that
allow effective and fine grained control and management over access
permissions.
The RFID Guardian~\cite{DBLP:conf/lisa/RiebackGCHT06} is a notable exception.
The main idea is to jam all reader to tag communication, except for reader
requests that satisfy a pre-defined privacy rule. This approach
has its own shortcomings. For one, it is extremely hard to ensure that all
reader to tag communication is effectively blocked in all cases. Moreover, tags
themselves are not protected at all, leaving them vulnerable when the Guardian
is out of range or malfunctioning. 

We refer to Juels~\cite{juels2006rfid-secpriv-survey} 
(and the excellent bibliography\footnote{\url{www.avoine.net/rfid/}}
maintained by Gildas Avoine)
for a much more extensive survey of proposed solutions,
and \cite{juels2007strong-privacy-rfid} for a more formal analysis of the
privacy properties actually achieved by some of the proposed
authentication protocols.

\subsection{On the hardware cost of cryptography}

We base our work on (relatively) new insights regarding the amount of hardware
required to implement symmetric key cryptosystems as compared to hash
functions. Traditionally, such hash functions are perceived to be the most
basic (and therefore most efficiently implementable) building blocks, and hence
have been used extensively in protocol designs for RFID. 
This is wrong.
In fact, the ECRYPT report on light weight
cryptography~\cite{oswald2006light-weight-crypto}
states
\begin{quote}
Current standards and state-of-the-art low power implementation techniques
favor the use of block ciphers like the AES instead of hash functions as the
cryptographic building blocks for secure RFID
protocols.
\end{quote}
Currently there is an AES-128 design with only 3k4 gates, 
which uses 3 $\mu$A at 100 kHz and 1.5 V in 0.35$\mu$ technology~\cite{feldhofer2005aes-grain-sand,poschmann2007rfid-crypto,feldhofer2007strong-crypto-rfid}. 
Maximum throughput is 9.9 Mbps (encryption) or 8.8 Mbps (decryption). 
A SHA-256 design~\cite{DBLP:conf/otm/FeldhoferR06,feldhofer2007strong-crypto-rfid}
that also has been targeted specifically for the low end 
results in 10k9 gates that has a maximum throughput of 22.5 Mpbs. 
Current consumption is 15.87 $\mu$A at 100 kHz and 3.3 V in
0.35$\mu$ technology. Other light weight symmetric cipher designs exist as
well~\cite{bogdanov-present}. 
The quoted AES design is 3 times as small (and thus also 3 times cheaper), 
and consumes less than 10\% of the power needed by the SHA-256 design, 
and is only about 2.5 times slower in terms of throughput.
We use these observations to design efficient protocols that incorporate
tag and reader authentication with session key establishment and 
fine grained control and management over access permissions.
Advances in reducing the hardware cost of implementing public cryptography have
also been made. Current implementations of (hyper)elliptic curve cryptography 
require 15,4k gates, executing one scalar multiplication in 243 ms when clocked
at 323 kHz~\cite{Fan2009HECC,DBLP:journals/tc/LeeSBV08}. Still the gate count
and the processing time are much higher than for symmetric cryptography, making
symmetric cryptography the preferred choice for lower cost tags.

\subsection{Our contribution}

Our contribution is to propose a model and corresponding protocols that
allow fine grained, effective and efficient control over access
permissions for RFID tags, that respect the privacy of the users. The model is
enforced by the tags themselves. The protocols use authentication as a basic
component, and we propose a novel combination of
(universal) re-encryption~\cite{juels2003squealing,golle2004reencryption}
with symmetric cryptography based authentication~\cite{ISO9798-2}
to obtain a reasonable level of privacy
protection without using public-key cryptography on the tag, and without the
need for the reader to start a time consuming key-search algorithm to
find the key to use for authentication. Although such key-search algorithms are
highly popular in the research community because of their superior privacy
properties, we believe they are unreasonable for large scale applications that
may involve millions of tags (and hence keys). 
Finally, our protocols are resistant to stolen
reader attacks, using techniques from~\cite{avoine2009compromised-readers}.
A detailed description of the properties of our authentication protocol is
presented in Sect.~\ref{sec-auth}.

The model is quite loosely based on the "Resurrecting Duckling" paradigm of Anderson
and Stajano~\cite{StaA99,stajano2000duckling-what-next}.
Our model is general enough to capture several RFID
use case scenarios, like supply chain management, ticketing and ambient
home intelligence. 

The essence of the model is that a potentially dynamic system of access
permissions is defined. We generalise the concept of an RFID tag, and view such
a tag as a container of several data objects on which a reader wishes to 
execute certain functions. 
Such an object implements a particular application or
service on a tag, like a customer loyalty program, or a supply chain management
program (where the on-chip object stores the identity of the physical object to
which the tag is 
attached). This extends the notion of an RFID tag containing just a unique
identifier to slightly smarter data container, resembling the technology used
for the new biometric passports~\cite{hoepman2006crossingborders}. We believe
that in the end, the idea of only storing a unique identifier on the tag and
storing all relevant data on the physical object attached to the tag in a
corresponding data record in a centralised database is going to prove too
limitative in the future. For example, for privacy reasons it is better to
require physical proximity to read the data on the tag instead of having
that data available in a database all the time.
We refer to Sect.~\ref{sec-usecases} for more examples supporting our model.

Whether the reader is allowed to execute the function
depends on two constraining factors:
\begin{enumerate}
\item whether the owner of the \emph{on-chip object} has
given the reader a permission to execute the particular 
function on the particular object, and 
\item whether the owner of the \emph{tag} allows
the reader to access the tag at all.
\end{enumerate}
The protocols (and the observation that the real challenge in RFID privacy lies
in allowing controlled use of the RFID tag even after the point-of-sale)
are inspired on the work by Engberg~\cite{engberg2004zeroknowledge-rfid}.
The first constraint is enforced using specially crafted permissions. 
The second constraint is enforced by the mutual reader-tag authentication
protocol. 
This research is part of the PEARL\footnote{%
	 \url{www.pearl-project.org}
}
(Privacy Enhanced Architecture for RFID Labels) project.

The paper is structured as follows. We first describe a few distinctive use
cases of RFID and their associated requirements in terms of functionality and
privacy. In Sect.~\ref{sec-model} we present our system model. We then
show how our model captures the essence of the use cases, in
Sect.~\ref{sec-analysis}. We then continue to implement this model using
data structures (Sect.~\ref{sec-datastruct}), an authentication and session
key establishment protocol (Sect.~\ref{sec-auth}) and subsequent
protocols (Sect.~\ref{sec-prot}). We analyse their security in
Sect.~\ref{sec-secanalysis} and present some conclusions and further research
in Sect.~\ref{sec-concl}.

\section{Use cases}
\label{sec-usecases}

This section describes three use-cases or scenarios, each of which focusing on
one or more issues that need to be addressed by our model. In the next section
we lay down our model, and then analyse how well the model captures the real
life situations sketched in these use cases.

\subsection{Scenario 1 - Supply Chain Management}

As our first scenario, we take the supply chain scenario from
\cite{engberg2004zeroknowledge-rfid}. In this scenario, we show the need for a
proper definition of tag ownership and the controlled transfer thereof.

In retail, RFID tags are attached to individual
products after they have been manufactured for the purpose of enhancing the
supply chain efficiency. Such tags contain an Electronic Product
Code (EPC) that not only identifies the type of product as the more classical
bar codes do, but also includes a number that is unique to the individual
product. This allows retailers to better keep track of their inventory and
counter theft attempts, and provide better customer service as well.

Often, for privacy reasons, these RFID tags are zapped, killed or removed from
products at the point of sale. This is unfortunate. 
Many interesting after-sale services are possible if the tag allows for a more dynamic and
fine grained control of access to it by its current owner.
For example, it would open the way for manufacturers and
retailers to store post-sale service data onto the tag such as the production
run, date of sale, warranty data, repair history etc., thus providing the them
with a possibility for providing more efficient en effective service to the
customer. 

When a product, such as a TV set, computer, refrigerator etc. 
needs to be repaired at some point in time, its warranty may be voided if it is
not repaired by qualified service
organisations.
Using the tags as introduced above, the manufacturer of 
equipment installs a service object into the tag associated with each of its
products. In this object, the manufacturer writes all sorts of service data
that it does not want to disclose to parties other than qualified service
organisations. Then, it provides read permissions for such service objects 
only to accredited service organisations.

Many objects change owner during the course of their life, consider for example
second-hand cars, electronic equipment, etc. When the objects change owner, so
must the attached tags. But any objects on the tag (like the service object
above) must remain on the tag and must keep their data (like the repairs
performed on the object).

We can accommodate for this scenario if the following
requirements are met.
\begin{itemize}
\fixlistspacing
\item A party is allowed to
  communicate with a specific RFID tag if and only if (a) 
  it is the owner of the tag or (b) it is explicitly granted permission by the
  owner of the tag.
\item A party is allowed to access a specific data (object) on the tag if
  (a) it owns the data (object) or (b) the data (object) owner has
  explicitly enabled that party to access the data (object) using 
  a specific method. In other words, access is assigned for individual method
  calls on the object.
\item A party that owns a tag must be capable of transferring this ownership to
  another party, without deleting or changing objects.
\end{itemize}
Note that a consequence of transferring ownership of a tag is that the set of parties
that were allowed to communicate with the tag changes drastically, as only the owner
or a party that has a permission issued by this owner, may communicate with the tag.

\subsection{Scenario 2 - Smart Tickets}

Event Services Company (ESC) is large company issuing an advanced, RFID based,
smart event ticketing solution. With the ESC RFID tag embedded in a wristband,
users can buy tickets for music events on line, flash them in their wristband,
and prove purchase of the tickets at the entry gates of the event. Security is
important: ticket fraud, and especially black market sales are rampant in the
ticketing business. Moreover, ESC does not want free riders to use its 
ticketing system: only event organisers with a contract can use it.

Tom owns such a wristband. In fact, the fashion watch he got for his birthday
last year contained such an ESC tag for free. He regularly buys tickets for
soccer matches and music festivals on line and appreciates the fast and paper
ticket less service. Recently, the gay gym he visits changed to the ESC
service, allowing members to book specific slots in the gym in advance. Tom
sees this as a huge advantage (the gym can be crowded at unpredictable times),
but is concerned about his privacy: he would rather not get his visits to the
gym and to the soccer matches get connected. Luckily, ESC was well aware of
these concerns when designing the system, and instead of storing the tickets on
a central server they are all stored on the ESC tag of the user.

Apart from the requirements from the first scenario, this adds the following
requirements.
\begin{itemize}
\fixlistspacing
\item Data related to the physical object is stored on the tag, and not
  (necessarily) on a central server.
\item Permission to install an object may be restricted depending on the application (in this case: the tag issuer).
\end{itemize}

\subsection{Scenario 3 - At the Hospital}

In our third scenario, we consider the situation where a doctor implants
a sensor tag into one of its hospitalized patients, e.g. by having him swallow a pill 
containing this sensor tag. In this scenario, we show the need for a proper 
definition of transferal of object ownership.
This scenario is inspired by the hospital scenario from~\cite{StaA99}.

Consider a patient whose heart condition, respiration and the like need 
to be monitored, and a high-tech monitoring device exists that acts like
a tag as in the previous scenario's. Because of price and the fact that they 
are not needed all that often, hospitals own such devices, and only a modest
amount of them.

Whenever a patient's condition is to be monitored, its doctor can decide to
implant such a device into the patient, e.g. by having him swallow a pill
containing the device. Within the body, the sensor starts monitoring the
patient's condition, filling an object that is specific for the sensor. Doing
so, a sensitive amount of personal data is gathered within the object, and it
is part of the doctor's job to ensure that privacy is preserved.

Since the doctor uses the sensor, he must have pretty much full
control. However, he must also be able to assign read permission to
e.g. nurses. This requires him to actually own the object. Note that he should
not own the device itself, as this would allow him to (dis)allow other parties
access to other parts of the device as well, which, if that results in a
catastrophe, will put the blame with the hospital.

We can accommodate for this by adding another requirement
\begin{itemize}
\fixlistspacing
\item A party that owns an object must be capable of transferring this
   ownership to another party. 
\end{itemize}

\section{System model}
\label{sec-model}

The system model describes the different entities in the system, their mutual
relationships, and the operations that they can perform on each other.

\subsection{Notation}

We use $\symkey$ to denote a symmetric key, possibly subscript\-ed to denote its
owner, and use $s$ to denote a symmetric session key. 
We use $\pubkey$ for a public key and $\key$ for the corresponding private key.
Hash functions are
denoted by $h(\cdot)$. We write $\xor$ for the exclusive-or operation, and
$\concat$ for concatenation of bit strings. $\enc{k}{m}$ denotes the encryption
of message $m$ with symmetric key $k$ using some symmetric cipher, typically
AES. $\mac{k}{m}$ denotes a message authentication code (MAC) for message
$m$ derived from
a symmetric cipher (for instance CMAC
\cite{NIST-PUB-800-38B,DBLP:conf/crypto/BlackR00}) using key $k$.
Finally, $\macenc{k}{m}$ denotes the authenticated encryption of $m$ with key
$k$, for instance by appending the MAC of the ciphertext 
\cite{DBLP:conf/asiacrypt/BellareN00}.

\subsection{Tags and readers}

A \term{tag} $\rfidtag$ is a piece of hardware that contains data. 
At the very minimum, tags store a bit string that can be read and sometimes written. 
Usually, tags store several values that can be grouped together as tuples 
because of their logical use. More complex, smart card like tags, contain
ISO 7816~\cite{ISO7816} like file structures. We assume that for the anti-collision protocol random identifiers are used (or else all bets to achieve some level of privacy are off).

We assume readers are at least on-line some of the time to obtain fresh data and keys from the central back office.

\subsubsection{Classes and objects on tags}

The system model follows the object oriented (OO) metaphor, so that tags are
said to contain \term{objects}, each of which is a group of bit strings whose
structure is defined by the \term{class} that it instantiates.  
We use $\obj \in \class$ to denote that object $\obj$ is an instantiation of 
class $\class$.
For every class, each tag contains at most one instantiating object.
Every class defines a set of \term{methods}, each of which specifies a kind of
operation that may take place on objects that instantiate that class. Simple
methods specify how to read or perhaps write values in a tuple of a certain
type stored on a particular tag.  More complex cases methods might invalidate a
ticket on a tag, or increase an electronic purse balance.
We write $\method \in \class$ for a method $\method$ that is defined for 
class $\class$.  
Every method is defined in precisely one class.
Access to a specific method is controlled in one of three ways:
\begin{itemize}
	\item the method can be called iff the tag is not owned;
	\item the method can be called if the user has an appropriate permission;
	\item the method can be called by the domain owning the class.
\end{itemize}
The OO metaphor can be applied both to the resource constrained case where 
a tag contains only an identifier or a tuple of values, and to the case where 
complex data structures are stored on a tag.

\subsubsection{The tag management class}

Every tag always contains one instance $\objmngrobj$ of the \emph{tag management class},
initially with default settings. 
The tag management class implements functions to manage tag access and ownership. 
This allows us to implement tag and class management operations in a similar way as methods on ordinary objects, thus simplifying the implementation. Details are provided in Sect.~\ref{sec-prot}.

\subsection{Domains and Principals}

We use the term \term{domain} to refer to a (legal) entity that is capable of bearing responsibilities. 
Thus, companies, organisations and governments are considered to be domains, 
as well as individual (adult) persons. 
We use the term \term{principal}, or \term{actor}, to refer to a resource 
(e.g. a person, or a running application within a computer) 
that is capable of acting on behalf of, c.q. under the responsibility of, a domain.
While a principal $\actor$ may act on behalf of different domains over time, 
and the change frequency thereof may be very high, 
we assume that at any particular point in time $\actor$ acts on behalf of precisely one domain $\domain$.
Note that in case of natural persons, who can both act as bear responsibility, the common practice where a single name is used to refer to the person both as an actor and as a domain, may cause considerable confusion. 
Thus, if a principal $\actor$ acts on behalf of a domain $\domain$ at a given point in time,
 then $\domain$ is responsible for everything that $\actor$ does at that time.
Since the domains bear the responsibilities, we have no compelling need to distinguish 
between the various principals that may act on behalf of a given domain, and thus 
we assume every domain to be inhabited by exactly one principal. 
We use $\domains$ to denote the set of all domains. 

\subsection{Ownership}

We use the term \term{owner(ship)} to refer to the responsibilities associated
with controlling tags, objects, etc.
Since responsibilities are born by domains, ownership can only be assigned to
domains. Ownership can be transferred by the owning domain to another
(accepting) domain.  

Thus, the \term{tag owner} for a tag $\rfidtag$ is a domain that bears the
responsibility for controlling access to $\rfidtag$, i.e. for issuing and
revoking the associated permissions. Also, it controls the permissions
associated with other tag related functionality, such as the creation of
objects or the transferal of tag ownership.  We use $\rfidtagowner$ to denote a
tag owner and $\rfidtagowners$ to denote the set of tag owners, so
$\rfidtagowner \in \rfidtagowners$ and $\rfidtagowners \subseteq \domains$.  We
write $\rfidtag \in \rfidtagowner$ to indicate that tag $\rfidtag$ is owned by
$\rfidtagowner$.

The \term{class owner} is responsible for
controlling access to objects that instantiate this class,
i.e. for issuing and revoking permissions for executing methods defined by that
class. 
We write $\class \in \classowner$ to mean that class $\class$ is owned
by domain $\classowner$ (i.e. its class owner).

Note that if a class owner $\classowner$ owns a class $\class$, then
(initially) it also owns every object $\obj \in \class$.  Thus, object
ownership is (initially) implied by class ownership.  However, ownership of
individual objects may be transferred to other domains later on. 
If that happens, the class owner is not necessarily the owner of all objects
of that class.

\subsection{Permissions}

Every \term{permission}, i.e. the right to access a tag or the right to execute
a method on an object, is issued by the domain that owns the tag or the
object. Also, permissions are issued to domains rather than to principals,
because domains can bear responsibilities associated with using such
permissions, which principals cannot. In our model, a permission that has been
issued to a domain can be used by any principal that acts under the
responsibility of that domain. Consequently, if misuse of a permission can be
traced back to the domain the permission was issued to, this domain can be held
accountable. 
It is outside the scope of this paper whether or not 
a domain limits the use of permissions that it has been assigned 
to a subset of the actors acting on its behalf, or sanctions misuse thereof.

One of our main contributions is the distinction we make between accessing
(i.e. communicating with) tags and accessing (i.e. executing methods on)
objects on a tag. A consequence of this distinction is that it requires two
rather than one permission to access an object on a tag: one permission is
needed for accessing the tag on which the object is stored (which is granted by
the tag owner), and the other
permission is required to execute the appropriate method on that object (which
is granted by the object owner). 
Moreover, these permissions are implemented quite differently (as described in
more detail in Sect.~\ref{ssec-permissions} and ~\ref{sec-auth}). The first permission
is checked using a mutual tag-reader authentication protocol, which verifies
that the reader domain occurs in a list of permitted domains. The second
permission is implemented using a permission token that encodes the permission
to access a particular method on an object. Thus,
manipulation of an object on a tag is controlled both by the tag owner and
object owner.

\subsection{Operations on a Tag}
\label{ssec-operations}

Operations are performed by actors (readers) acting on behalf of a domain.
Operations can only be performed when the actor acts on behalf of a domain that
has permission to do so. While other operations are certainly conceivable, 
we consider only the limited set of basic operations as specified in Sect.~\ref{sec-prot}.

The most basic operation the model must support is calling a method on an
object of a certain class stored on a particular tag. For this, two permissions are
required: first, the domain must be allowed to access the tag, and secondly the domain 
must be allowed to execute the method on (the class of) the object. Note that access to a
method is initially granted at the class level. So access rights for a
particular method initially apply to all objects of that class.

The creation of permissions is done off-tag, as is the distribution 
thereof\footnote{%
	The word 'capability' might be more appropriate than the word
	'permission'.
}.
Tag ownership is controlled through the following functions:
\begin{itemize}
\fixlistspacing
\item \func{TakeTagOwnership}: Set a specific domain as the tag's owner. Can be
  executed by any domain as long as the tag is not owned.
\item \func{TransferTagOwnership}: Transfer ownership of a tag from its tag
  owner to another domain. Can only be executed by the owner of the tag.
\item \func{RelinquishTagOwnership}: Relinquish ownership of a tag so that the
  tag is no longer owned. Can only be executed by the owner of the tag.
\end{itemize}
Tag access is controlled through the following functions:
\begin{itemize} 
\fixlistspacing
\item \func{GrantTagAccess}: Allow a specific domain to access a tag. 
\item \func{RevokeTagAccess}: Disallow a specific domain to access a tag.
\end{itemize}
These functions are only executable by the tag owner.


Object management is controlled through the following functions:
\begin{itemize} 
\fixlistspacing
\item \func{InstallObject}: Create an object and set the class key. Can only be executed by the tag owner, or any domain with a permission issued by the tag owner.
\item \func{UpdateObject}: Overwrite the contents and the code of an object. Can only be executed by the class owner, or any domain with a permission issued by the class owner.
\item \func{UpdateClassKey}: Change the class key associated with an object. Can only be executed by the class owner. This function can (also) be used to transfer ownership of objects.
\item \func{DeleteObject}: Destroy an object and its associated class key. Can only be executed by the class owner, the tag owner, or any domain with an appropriate permission issued by the class owner.
\end{itemize}
As said before, this paper only describes a basic set of operations that will
allow us to implement the scenarios from Sect.~\ref{sec-usecases}.
Other operations are certainly possible and can easily be added to the model
and implemented in a similar fashion as the basic operations.

\section{Analysis}
\label{sec-analysis}

The system model from Sect.~\ref{sec-model} should allow us to implement a
large set of common privacy friendly uses of RFID technology. To capture these
use cases, we sketch\-ed three different scenarios in Sect.~\ref{sec-usecases}.  
We now briefly verify that our model indeed allows us to implement these three
scenarios. The security and privacy properties are analysed after
we have presented the protocols that implement the operations -- they do not
depend on the model, but on the actual implementation.

\subsection{Mapping of Scenario 1}


Product tags that comply with our model would be attached to the product when
manufactured. For every product type that a manufacturer M produces, M defines
an object class \objname{Service} that contains data and access methods that is
relevant to the manufacturer, including a.o. production data, production plant,
serial numbers and so on. First, M takes an unowned tag and takes ownership
thereof (executing the tag's function $\func{TakeTagOwnership})$. 
Tag owners can then execute \func{InstallObject}, which is what M uses to create
the \objname{Service} object on the tag.
For each of the methods on this object, 
M creates permissions (see Sect.~\ref{ssec-permissions}) that M assigns to
itself so that it can access all methods itself.
Note that M only needs to create such permissions once, as they will be
usable on every \objname{Service} object M creates.

To accommodate the service-organisation scenario, all that M needs to do is
create a read-permission for \objname{Service} objects for every organisation
that it has accredited for servicing M's TV sets, and send this permission to
the appropriate organisation in a secure manner. This way, only accredited
organisations (and M) may read \objname{Service} objects.
Note that service organisations cannot yet read the
\objname{Service} objects since they do not have permission to access the tag
itself. This is done later when the consumer becomes the tag owner.

Whenever a retailer R sends M an order for a number of TV sets, M prepares the
delivery. For every tag in this delivery, M first writes appropriate data into
the \objname{Service} object so that it says to which retailer it will be delivered, as
well as other information M might later need. Then, M transfers ownership of
the tag to R (\func{Transfer\-TagOwnership}, which means that M no longer is
capable of accessing the \objname{Service} object because it can no longer access the tag
(see Sect.~\ref{ssec-transferownership} for details). Still, M's service
object remains on the tag and all permissions that it has issued to itself and
the accredited service organisations remain valid. 
Then, M sends some data to R in a sufficiently secure manner, thus enabling R 
to gain ownership of the tags (See Sect.~\ref{ssec-transferownership}). 
While the shipment is in transit, only M and R can take control of it as they
have the data to regain ownership. For anyone else, the tag is useless as they
cannot communicate with it. 

For use in its retail processes, R has already defined an object class \objname{Retail},
and like M, R has created and distributed permissions for \objname{Retail}
objects to itself, and other domains as necessary. 
Thus, when R receives the data that M has sent as well as the shipment, 
R can take control of each tag, and create a \objname{Retail} object on each 
of them, filling it with data relevant to R's retail process. 
Note that the tags still contain \objname{Service} objects, but R can only 
access such objects if it has been issued appropriate access permissions, i.e. if 
R is a service organisation that M has accredited.
Also note that R controls whether or not M can access its own service object, 
as M needs tag-access permissions which R can grant (\func{GrantTagAccess}) or 
deny (e.g. revoke using \func{RevokeTagAccess}).

When a customer C buys a TV set, R updates its service object and subsequently
transfers ownership of the tag to C. C may subsequently grant R and M access to
the tag, that would allow them to work (only!) with their own service
objects (and objects for which they have been issued a permission by the corresponding 
class owner). Also, C can install a data object of its own on the tag provided an
appropriate class has been defined and permissions created. Also, C can resell
the TV set to C' and simply transferownership of the tag to C'. 

If C is not interested in managing the tag, then R may safely keep tag ownership 
as no other domain than R (and perhaps M) would be able to use the tag, 
and still then only using their own data. 
A more difficult situation is if C had taken up tag ownership, but sells it to 
a party C' that is not interested in taking tag ownership. While we think there
may be several solutions here, we leave this case outside the scope of this paper.
Thus, throughout the lifetime of a tag, each owner M, R, or C has full control
over who can use the tag and who cannot. Also, M, R, or C can install their own
data the confidentiality of which is under their own control.

\subsection{Mapping of scenario 2}
In this scenario, ESC (Event Services Company) is the first to take ownership
of the tag using \func{TakeTagOwnership}. Using the default (known) class key
of the tag management object $\objmngrobj$, it creates a permission to call 
\func{UpdateClassKey} to set the class key of $\objmngrobj$ to its own 
secret value. This key is used to create permissions for every event 
organisation that has a contract with ESC, to install objects through
$\objmngrobj$ using \func{InstallObject}.

ESC transfers ownerships to consumers buying ESC tags using
\func{Transfer\-Tag\-Ownership}. Now users buying tickets from certain organisations
first grant access to the tag for these organisations through
\func{GrantTagAccess}. These organisations then install their own ticket object
calling \func{InstallObject} with the relevant ticket data on the tag. They
need permission from ESC (as described in the previous paragraph) to do so.

\subsection{Mapping of Scenario 3}

With our model, we can show how in scenario 3 ownership of objects can be
transferred between parties. 

We start out with a doctor D that works at a hospital H which has a patient P
and a nurse N. H owns a high-tech monitoring tag (device) T, which contains at
least one object being of the class Tmon which has methods implementing all
sorts of monitoring functions. 

When D decides to implant T into P, D becomes responsible for the use of
functions of the Tmon object. While it is undesirable to transfer ownership of
T to D, it is desirable to transfer ownership of the Tmon object from H to D
because this allows D to control who may use which function of the Tmon
object. Thus, when D borrows T from H, H transfers ownership of the Tmon object
to D (issuing a permission to D to call \func{UpdateClassKey} on Tmon). 
This immediately makes all
existing permissions obsolete that H has assigned to any domain for this
particular Tmon object. However, such permissions remain valid for all Tmon
objects that H still owns. 

Now, D can issue permissions to the Tmon object, e.g. to nurse N that nurses
the patient. 

When P is dismissed from the hospital, T is removed from P, and ownership of
the Tmon object is returned to the hospital. This immediately invalidates the
permission that N has for the Tmon object. However, as long as the validity
period of this permission has not expired, N can still use it to access Tmon
objects on other tags (provided N has access to the tag (which is controlled by
the hospital) and the Tmon object is owned by D. 

\section{Data structures}
\label{sec-datastruct}

In this section we describe the data structures stored by the tags, and the keys
and permissions used by the domains to access the data on a tag. In the next
section we describe the implementations of the operations that can be performed
on a tag.

\subsection{Keys}

To implement permissions, the system uses the 
the following types of keys. Some keys (the domain key pairs
$\KD$, $\kD$) are asymmetric keys, the other keys are symmetric keys.
\begin{description}
\fixlistspacing
\item[Tag access keys $\kta$.] 
  Access to tags is controlled using tag access keys $\kta$. These keys are
  unique to a tag, and derived from the tag identifier $t$ using a master
  access key $\kMA$ through key diversification~\cite{AndB96}
  by $\kta = \enc{\kMA}{t}$.
\item[Master access keys $\kMA$.]
  Each domain has a master access key $\kMA$.
  Readers in a domain use this master access key $\kMA$ to derive tag access 
  keys from tag identifiers. Each tag thus stores, for each domain that is
  allowed to access it, a different tag access key. 
\item[Domain key pairs $\KD$, $\kD$.] 
  Each domain keeps a unique ElGamal public/pri\-vate domain key pair $\KD$,
  $\kD$. 
  These keys are used in the
  authentication protocol to preserve privacy of the tag identifier $t$.
  To thwart stolen reader attacks, readers get a new pair of keys every
  once in a while. These keys are stored in the array $\eka[]$.
\item[Class keys $\kc$.] 
  For each class there exists a unique class key $\kc$.
  The class key is used to encode access permissions to the class methods.
  A tag stores, for each object, the corresponding class key
  to verify such permissions.
  Class owners know all the class keys of the classes they own.
  Changing the class key of an individual object can be utilised to transfer
  ownership of that particular object. Conceptually, however, this makes the
  object member of another class (albeit with the same structure and methods
  as the class it originally was a member of).
\end{description}

\subsection{Other data stored on the tag}

A tag $\rfidtag$ also performs a bit of bookkeeping. Firstly, it records
a time stamp $\now_\rfidtag$ that approximates the current date and time (see
below), initially $-\infty$.  
Tags also store several objects, each of a class $\class$ together
with the key\footnote{%
	This is a weakness that seems to be unavoidable: the owner of
	the tag can in principle recover the class key; the install
	procedure should ensure that the key cannot be captured in transit.
}
$\kc$. 
Also, a tag $\rfidtag$ keeps an access set $\accessset_\rfidtag$ that
stores, for each domain $\domain$ that is granted access to the tag, the
following three items.
\begin{itemize}
\fixlistspacing
\item An encrypted tag identifier $\encid$, equal to the ElGamal encryption
$(t \cdot \KD^x,g^x)$ of the tag identifier $t$.
\item The epoch $\epoch$ in which the encrypted tag identifier was created
  (for explanation see Sect.~\ref{sec-auth}).
\item The diversified tag access key $\kta$, which equals $\enc{\kMA}{t}$ 
for the master key $\kMA$ used by domain $\domain$. 
\item A boolean flag indicating whether this domain is the owner of the tag.
\end{itemize}
We interpret the access set as a dictionary indexed by domains
(as a domain can have at most one such tuple in the access set), and
write $\accessset_\rfidtag[\domain] = (\encid,\kta,b)$.
There is at most one domain that is the owner of the tag. We write
$\owner_\rfidtag$ for that domain (which equals $\bot$ if the tag is not owned
by a domain). 
Initially, $\accessset_\rfidtag = \emptyset$.

Finally, the tag stores the current session key $s$, which initially and
in between sessions equals a default value (denoted $\bot$, but which actually
is a valid key), and which is set to a
certain value as the result of a successful mutual authentication (in which case
the authenticated reader holds the same session key). It also stores the
domain of the reader that was authenticated in $\curdomain$ (which equals
$\bot$ in between sessions).

We usually omit the subscript from $\now$, $\owner$ and
$\accessset$.

\subsection{Permissions}
\label{ssec-permissions}

To grant a domain $\domain$ access to a method $\method$ on an object of
class $\class$ up to time $\maxtime$, the class owner $\classowner$ 
generates a \term{permission token}
\[
\pcfod = \enc{\kc}{\method,\domain,\maxtime} 
\] 
and sends this to the domain $\domain$. This permission token expires as soon as the
current 
time exceeds $\maxtime$. Tags use $\now$ as their estimate of the current time
to verify this. 
This is updated after each successful call of a method on the tag (which
includes the current time as asserted by the caller).
It is also set to the current time when the first domain takes ownership of the
tag. 
A similar method is also used by the European RFID
passports~\cite{bsi2006extendedaccesscontrol,hoepman2006crossingborders}.

\section{Mutual authentication and establishing a session key}
\label{sec-auth}

A basic step underlying the protocols that implement the operations that access
a tag, is to mutually authenticate a tag and a reader, and to establish a
session key among them\footnote{%
	Actually, from a privacy perspective, we are only interested in
	authenticating the \emph{reader}. Only after the reader is
	proven  authentic, and has permission to access the tag, the tag
	has to identify and authenticate itself. However, since we are unable
	to use public key cryptography on the tag, and we do not wish to store
	global shared secrets on the tag, we are left with using key
	diversification based on the identity of the tag. Hence
	authenticating the reader as well as the tag simultaneously 
	seems to be the only way forward. 
}. 

Below we present a protocol that is efficient for both the reader and the tag. 
In principle it combines elements of three different known authentication
protocols to strike a balance between tag and reader efficiency, achieve a
robustness against a reasonably large class of adversaries, and achieve a
certain level of privacy as well. In fact it combines a standard, 
ISO/IEC 9798-2~\cite{ISO9798-2} based symmetric key authentication protocol, 
with (universal) re-encryption~\cite{juels2003squealing,golle2004reencryption}
to avoid the costly key search, and 
a counter based approach to invalidate keys from stolen 
readers~\cite{avoine2009compromised-readers}.
To further enhance privacy, users may perform a separate re-encryp\-tion of all
identifiers on a tag at any time.

To be precise, the protocol achieves the following properties
\begin{description}
\item[mutual authentication]
  The reader and the tag are mutually authenticated.
\item[soft privacy]
  Tags can only be traced in between two successful
  re-encryptions (including the re-encryp\-tion performed during an
  authentication). Except for the reader performing the 
  re-encryption, no other reader or eavesdropper can link the presence of the 
  tag after the re-encryption with an observation of this tag before the
  re-encryption. 
\item[owner-controlled privacy]
  Tag owners can re-encrypt all tag identifiers for all domains at any time on the tags they own.
\item[resilience to tag compromise]
  Tags do not contain global secrets. Hence a tag compromise does not affect 
  any other tags in the system.
\item[resilience to reader compromise]
  Stolen readers (or otherwise compromised readers) will not be able to 
  recognise or access tags, once those tags have been in contact with another
  valid reader after the compromise~\cite{avoine2009compromised-readers}.
  A similar property is achieved by the European biometric
  passports~\cite{bsi2006extendedaccesscontrol,hoepman2006crossingborders}. 
\item[reader efficiency]
  The reader performs a constant number of operations.
\item[tag efficiency]
  The tag performs only a constant number of symmetric key cryptography
  operations. 
\end{description}

\begin{figure*}[tp]
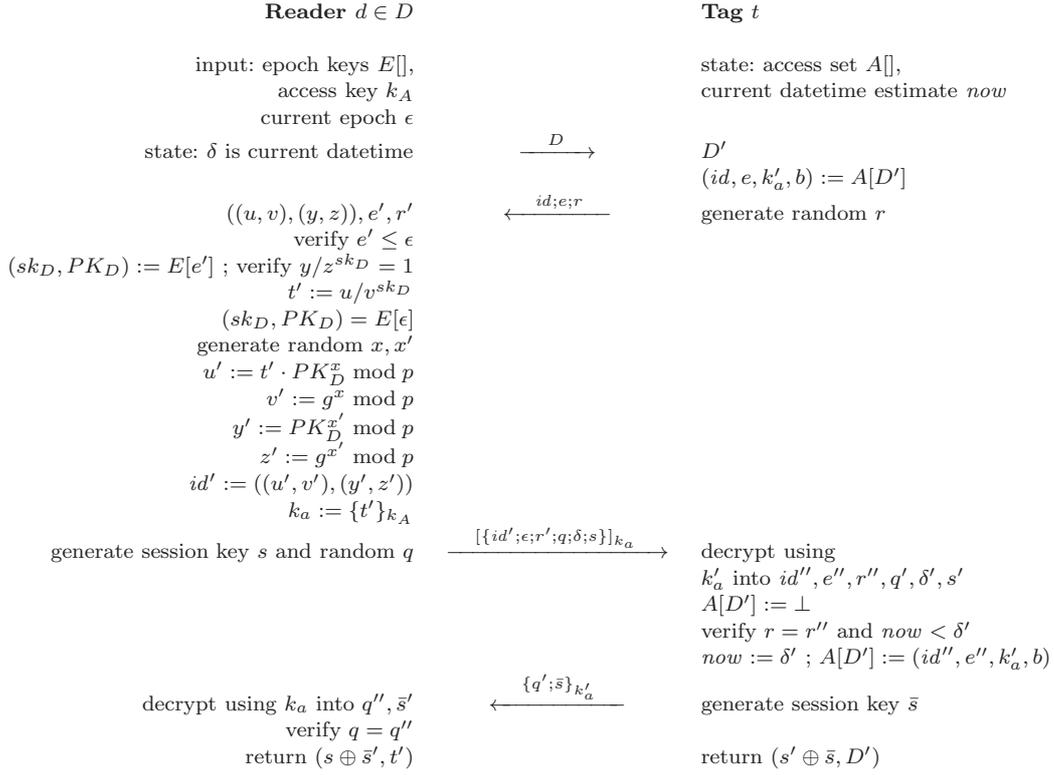

\begin{center}
\begin{tabular}{rcl}
\textbf{Reader $\actor \in \domain$} & & \textbf{Tag $\rfidtag$} \\
 & \\
input: epoch keys  $\eka[]$,
		&	& state: access set $\accessset[]$, \\
access key $\kMA$ 
		&	& current datetime estimate $\now$ \\

current epoch $\curepoch$ \\
state: $\datetime$ is current datetime 
		& \sendright{\domain}
			& $\domain'$ \\
		&	& $(\encid,\epoch,\kta',b) \assign \accessset[\domain']$ \\
$((u,v),(y,z)),\epoch',r'$
		& \sendleft{\encid \concat \epoch \concat r}
			& generate random $r$ \\
verify $\epoch' \le \curepoch$ \\
$(\kD,\KD) \assign \eka[\epoch']$ ;
verify $y / z^\kD = 1$ \\
$\rfidtag' \assign u / v^\kD$ \\
$(\kD,\KD) = \eka[\curepoch]$ \\
generate random $x,x'$ \\
$u' \assign \rfidtag' \cdot \KD^{x} \bmod p$ \\
$v' \assign g^{x} \bmod p$ \\
$y' \assign \KD^{x'} \bmod p$ \\
$z' \assign g^{x'} \bmod p$ \\
$\encid' \assign ((u',v'),(y',z'))$ \\ 
$\kta \assign \enc{\kMA}{\rfidtag'}$ \\
generate session key $s$ and random $q$
	 	& \sendright{\macenc{\kta}{\encid'\concat \curepoch \concat r' \concat q \concat \datetime \concat s}} 
			& decrypt using \\
		&	& $\kta'$ into $\encid'',\epoch'',r'',q',\datetime',s'$ \\
                &       & $\accessset[\domain'] \assign \bot$ \\
		&	& verify $r = r''$ and $\now < \datetime'$ \\
		&	& $\now \assign \datetime'$ ;
		          $\accessset[\domain'] \assign        
                          (\encid'',\epoch'',\kta',b)$ \\
decrypt 	
using $\kta$ into $q'', \bar{s}'$ 
		& \sendleft{\enc{\kta'}{q' \concat \bar{s}}} 
			& generate session key $\bar{s}$ \\
verify $q = q''$ \\	
return $(s \xor \bar{s}',\rfidtag')$
		&	& return $(s' \xor \bar{s},\domain')$ 
\end{tabular}
\end{center}
\caption{Authentication and session key agreement.}
\label{fig-protenc-auth-2}
\end{figure*}

The protocol we present below explicitly checks the correctness of the 
responses, that
may contain additional information for that purpose, to positively authenticate
the other party. Another option is to rely on implicit
authentication through the session key that is established as well: if the
authentication fails, both parties will have different values for the session
key, and therefore subsequent protocol steps will fail.

Note that in the description of the protocols we do not explicitly describe the
behaviour of a principal if it detects such an error. Instead we use the
convention that if an internal check fails, the principal continues to send the
expected messages at the appropriate times, with the appropriate message
format, but with random message content. This is necessary to preserve privacy,
as observed by 
Juels~\etal~\cite{juels2006strong-privacy-rfid,juels2007strong-privacy-rfid}.

Our protocol (see Fig.~\ref{fig-protenc-auth-2})
is an extension of the the ISO/IEC 9798-2~\cite{ISO9798-2}
standard, using diversified keys~\cite{AndB96} to avoid sharing keys over many
tags\footnote{%
	The first encrypted message is also protected by a MAC, because
	the contents of the message should not malleable while keeping the
	response to the challenge intact. This is not guaranteed if one only
	encrypts the message, \eg in ECB mode.
}. 
The tag stores such a diversified
tag access key $\kta'$ that corresponds to
$\enc{\kMA}{\rfidtag}$. 
To compute this diversified key from the master access key
$\kMA$ it stores, the reader needs to
learn the tag identifier $\rfidtag$. 
This cannot be sent in the clear for privacy reasons. The solution is 
to encrypt the tag identifier $t$ against the public key of the reader
to obtain $\encid$, and let the reader re-encrypt~\cite{juels2003squealing}
that value with every authentication run. This way the tag does not have to perform any public key operations.
Note that the re-encrypted value is only used as the new tag identifier after a
successful authentication of the reader. This avoids denial-of-service attacks.
Finally, the re-encryption keys stored by the readers are updated every time
a reader is stolen. Every time this happens, a new \emph{epoch} is
started.
Stolen readers no longer receive keys for future epochs. Tags that 
authenticate successfully, receive a new encrypted identity, encrypted against
the most recent epoch key. This makes it impossible for compromised readers to
track this tag.

Note that corrupt readers can update the tag identifier to an arbitrary
value. If that value is not recognised as a tag identifier by a genuine reader
in a next authentication run, this reader will send random data to the tag. The
tag will detect this and set 
$\accessset[\domain] \assign \bot$. The tag will then stop responding to
requests from this domain. 
Without this countermeasure, the arbitrary value for the identifier would never
change and the tag would be traceable forever.

The protocol can be extended using ideas from Hoepman~\etal\cite{Hoe00} by
storing so called authentication credit on the readers, that cannot be used to
generate valid authentication responses. This way, readers do not need to store
master keys, and therefore need to be less trust\-ed, or can be operated in less
trusted environments.

At the reader side the protocol returns the tag identifier and the
session key to be used.
For a call to such an
authentication protocol run in the protocols below
we write
$\id{AuthenticateR}(\kD,\KD,\kMA)$.
At the tag side, the protocol returns the session key, as well as 
the authenticated domain.
We write $\id{AuthenticateT}()$ for this call.

\subsection{Re-encryption}
\label{ssec-reencrypt}

The protocol uses re-encryption, or rather \emph{universal}
re-encryption~\cite{golle2004reencryption}, as follows. 
We use the ElGamal encryption scheme~\cite{DBLP:journals/tit/Elgamal85}
over a cyclic group $G$ of order $q$.
To be
concrete, and to achieve semantic security~\cite{DBLP:conf/pkc/TsiounisY98}, we
choose two primes $p$ and $q$ such that $q \| (p-1)$ (\ie $q$ is a divisor of
$(p-1)$) and choose as $G$ the cyclic subgroup of $\Z_p$ with order $q$, and 
pick a generator $g$ for $G$. These are global, system wide, constants.

Each domain has, for each epoch, its own public/pri\-vate key pair ($\KD,\kD)$
where $\kD$ is a random integer between $1$ and $q-1$, and $\KD = g^\kD$.
The tag identifier $t$ is encrypted, using ElGamal, as
\[
(u,v) = (t \cdot \KD^x, g^x)~,
\]
where $x$ is a random value in $[0,q-1]$. 
To allow re-encryption by readers that do not know the corresponding private
key, each tag stores with each encrypted tag identifier a corresponding
re-encryption factor
\[
(y,z) =  (\KD^{x'}, g^{x'})~,
\]
where $x'$ is a new random value in $[0,q-1]$. 
Note that this is basically an encryption of
the value $1$ against the same key. Because ElGamal enjoys the homomorphic
property that the multiplication of the encryption of two ciphertexts equals
the encryption of the multiplication of the corresponding plaintexts, we see
that $(uy,vz)$ in fact equals the encryption of tag identifier $t$.
The encrypted identifier now becomes
\[
\encid = ((u,v),(y,z))~.
\]

Readers store the key pairs for the epochs in an array $\eka[]$, storing
the keys for epoch $\epoch$ at $\eka[\epoch]$. This array is filled with epoch
keys up to and including the current epoch $\curepoch$, and grows in size over
time. 

To re-encrypt, a reader that knows the corresponding, most recent public epoch
key $\KD$ does the following. It generates new random
values $a$ and $a'$ in $[0,q-1]$ and computes
\[
(u', v') = (t \cdot \KD^{a} , g^{a})
\]
and
\[
(y',z') = (\KD^{a'} , g^{a'})
\]
and sends
\[
\encid'= ((u',v'),(y',z')) 
\]
to the tag.
Readers that do not know the current epoch key can use the re-encryption 
factor to compute a new encrypted identifier as follows. Again two random
factors $a$ and $a'$ in $[0,q-1]$ are generated, and then the reader computes
\[
(u', v') = (u \cdot y^a, v \cdot z^a )
\]
and
\[
(y',z') = (y^{a'} , z^{a'})
\]
and again sends
\[
\encid'= ((u',v'),(y',z')) 
\]
to the tag.

Requests to re-encrypt other encrypted tag identifiers can be issued by
authorised readers to the tag management object, see
Sect.~\ref{ssec-reencrypt-method}.
Typically, readers that are owned and operated by a tag owner will have the
permission to perform such re-encryptions. This way, owners of tags have
control over how easily their tags can be traced. 
Without universal re-encryption, only readers knowing the public key of the
domain can re-encrypt. If a tag is hardly ever accessed by such a reader
(consider for example a supply chain tag attached to a piece of clothing that
is never accessed again after the point of sale), such a tag is principle
unlimitedly traceable. By frequently re-encrypting
their tags, users can make such tags much less traceable. 

To decrypt, one simply verifies that $y/z^\kD = 1$ and
computes $u/v^\kD$, using the appropriate epoch key
stored in $\eka[]$. To avoid the need to search for the right key, the tag
sends, together with is encrypted identifier, the epoch in which it was last
updated\footnote{%
  This impacts privacy, in particular it allows one to trace 
  tags that are infrequently used and hence broadcast old epoch numbers.
  However, in the current protocol that is not a separate concern, as the same
  tag will broadcast the same encrypted tag identifier until it is successfully 
  updated (in which instance its epoch will be set to the most recent epoch,
  which contains a large number of tags).}.

\subsection{Alternative approaches}

In the course of developing the above algorithm, we have considered various
alternatives. The main drawback of the above protocol is that tags are
traceable in between re-encryptions. Every malicious reader that claims to be
from domain $\domain$ will receive the current encryption of the
identifier. This can be solved in two ways, both incurring another, more
severe, drawback.

The first option is to let the tag (instead of the reader) do the re-encryption
each time it is queried by a reader. Then the tag is no longer dependent on a
reader to provide it with a proper re-encryption, and malicious readers no longer
pose a threat. But this requires that the tag is capable of performing modular
exponentiation at reasonable speed. This is out of scope for low cost
tags. Moreover, if the tag can do that, then one might as well use an
authentication protocol using asymmetric cryptography. Such a protocol would be
much simpler than our current proposal.

The second option is to stop responding to requests from domain $\domain$ after
a fixed number of times, unless one such request was a full run of the
authentication protocol that updated the current encryption of the
identifier. This limits the time a tag can be traced, but
makes the system vulnerable to denial of service attacks.

Finally, we considered another approach where the tag would randomly encrypt
its tag identifier to a symmetric domain key $\symkD$, sending
\[
\enc{\symkD}{r,\rfidtag}
\]
to the reader at the start of the authentication protocol\footnote{%
	This message should \emph{not} be encrypted in ECB mode, but in CBC
	mode (if the nonce and the tag identifier together do not fit
	inside a single block). The point is that the random value $r$
	preceding the tag identifier should randomise the encryption of the
	\emph{whole} message, in particular the encryption of $\rfidtag$, to  
	preserve the privacy of the tag.
}. By including the
random $r$, the whole message is randomised, and tags become
untraceable. However, $\symkD$ is stored on all tags accessible by domain
$\domain$. Because tags are not tamper proof, this key is not protected and will
become known after some time. From that time on, these tags become traceable
and privacy is lost.

\section{Protocols}
\label{sec-prot}

Below we will describe protocols that implement the operations
from Sect.~\ref{ssec-operations}. We take a rather generic approach. Instead of
implementing special protocols for each of these operations, we
in fact model all these operations
either as calls on normal objects (\func{Delete\-Object} and
\func{UpdateObject}), or as  
special methods of the tag management object $\objmngrobj$ (all other
operations).
That is, we present pseudocode for the body of each of these operations as if
they were methods of a certain object, operating on the state of the object and
or operating on the state of the tag. 

This way, the only 'protocol' that we need to
describe now is how to securely call a method on an object stored on a tag. In
fact, this protocol is split in three sub-protocols. The first sets up a
session and a shared session key, the second securely calls the method using
the session key to secure the channel and using permission tokens to verify the
legitimacy of the request, and the third closes the session.

Note (\cf Sect.~\ref{sec-auth}) again that we do not explicitly describe the
behaviour of a principal if it detects an error.

\subsection{Calling a method}
\label{ssec-callingamethod}

\begin{figure}[t]
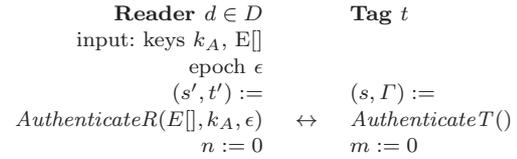

\begin{center}
\begin{tabular}{rcl}
\textbf{Reader $\actor \in \domain$} & & \textbf{Tag $\rfidtag$} \\
input: keys $\kMA$, E[] \\
epoch $\curepoch$ \\

$(s',\rfidtag') \assign $ 
	& 
		& $(s,\curdomain) \assign$ \\
$\id{AuthenticateR}(E[],\kMA,\curepoch)$
                & $\leftrightarrow$
                        & $\id{AuthenticateT}()$\\
$n \assign 0$
		&	& $m \assign 0$ \\
\end{tabular}
\end{center}
\caption{Setting up a session.}
\label{fig-protstartsession}
\end{figure}

To call a method $\method$ on an class $\class$, the reader $\actor$
belonging to
domain $\domain$ and the tag $\rfidtag$ first set up a session using
the protocol in
Fig.~\ref{fig-protstartsession}. 
This is nothing more than starting the
authentication protocol from Fig.~\ref{fig-protenc-auth-2}.
If this is successful, the reader and the tag share the same session key. 
Both initialise their message sequence counter ($m$ and $n$) to $0$. 

The actual method call follows the protocol in Fig.~\ref{fig-protcall}. This
protocol can be executed several times in a row, to execute several methods
within a single session.
Each message includes the current value of the message counter, and each
message is encrypted and MAC-ed with the session key. 
The message counters are incremented with every subsequent message within a session.
The receiver verifies the
included message counter to prevent replay attacks.

For each method call, the reader sends the corresponding permission token, which is
verified by the tag using the class key $\kc'$ of the class whose method is
called. It also verifies whether the permission token is still valid, using its own
estimate of the current time $\now$, and whether the permission token is bound to the
domain that was authenticated in the first phase.
Then the reader sends the method call
parameters, and the tag responds with the method result. If the method is
supposed to return no result, a random value is sent instead.
Note that the method is called with the name of the calling domain as the first
parameter.

To call a method on an object for which no permission tokens are necessary (which is
the case for some of the methods of the tag management object, see below), basically the same
protocol is used. In this case however, the caller does not have to send a
permission token, and the tag only verifies that the requested method on that object
is indeed callable without permission.

\begin{figure}[t]
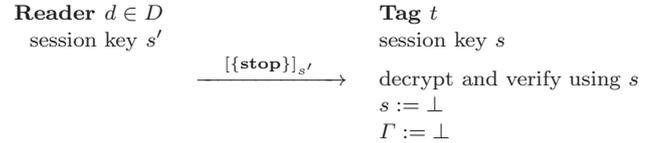

\begin{center}
\begin{tabular}{rcl}
\textbf{Reader $\actor \in \domain$} & & \textbf{Tag $\rfidtag$} \\
session key $s'$	&	& session key $s$ \\

 		& \sendright{\macenc{s'}{\mathbf{stop}}}
			& decrypt and verify using $s$  \\
		&	& $s \assign \bot$ \\
		&	& $\curdomain \assign \bot$ \\
\end{tabular}
\end{center}
\caption{Closing a session.}
\label{fig-protstopsession}
\end{figure}

Finally, to close a session, the protocol in Fig.~\ref{fig-protstopsession} is
executed.

\begin{figure*}[t]
\begin{center}
\begin{tabular}{rcl}
\textbf{Reader $\actor \in \domain$} & & \textbf{Tag $\rfidtag$} \\
session key $s'$	&	& session key $s$ \\
permission token $\pcfod$ 	&	& calling domain $\curdomain$\\
counter $n$		&	& counter $m$ \\

$p \assign \pcfod$ 
 		& \sendright{\macenc{s'}{n \concat \class \concat \method \concat \maxtime \concat p}}
			& decrypt and verify using $s$  \\
		&	& into $n',\class',\method',\maxtime',p'$ \\
		&	& verify $\now < \maxtime'$ \\
		&	& verify $n'=m$ \\
		&	& look up object of class $\class'$\\
		&	& and keep $\kc'$\\
		&	& verify $p' = \enc{\kc'}{\method',\curdomain,\maxtime'}$\\
		& \sendright{\macenc{s'}{n+1 \concat \mathit{parameters}}} 
			& decrypt and verify using $s$ into $n',x$ \\
		&	& verify $n'=m+1$ \\
decrypt and verify using $s'$ 
		& \sendleft{\macenc{s}{m+2 \concat \mathit{result}}} 
			& execute $f(\curdomain,x)$ \\
into $m',r$ \\
verify $m'=n+2$ 	& \\
$n \assign n+3$
		&	& $m \assign m+3$ 
\end{tabular}
\end{center}
\caption{Calling method $\method$ on class $\class$ using permission token $\pcfod$ valid until $\maxtime$.}
\label{fig-protcall}
\end{figure*}

\subsection{Tag ownership functions}
\label{ssec-transferownership}

The following methods on the tag management object $\objmngrobj$ 
implement transfer of ownership.
To relinquish ownership of a tag, the tag owner can execute
the following method.
\begin{description}
\item[\func{RelinquishTagOwnership}($\caller$)]:\\
	verify $\owner = \caller$ ; \\
	$\accessset \assign \emptyset$ (hence $\owner = \bot$)\footnote{%
		If so desired, resetting of $\accessset$ can be
		skipped. However, in that case the owner flag for $\Gamma$
		must be reset.
} ; \\
  $s \assign \bot$.
\end{description}
The functionality of \func{RelinquishTagOwnership} may be extended to include
the deletion of all objects (other than the tag management object), and the
resetting of information in the tag management object. 

To become the owner of an unowned tag, a domain calls the following method
\begin{description}
\item[\func{TakeTagOwnership}($\caller,\domain,\encid,\kta$)]:\\
	verify $\owner = \bot$ ; \\
	$\accessset[\domain] \assign (\encid,\kta, \id{true})$ ; \\
\end{description}
where the caller of $\func{TakeTagOwnership}$ from domain $\domain$ 
has received the tag identifier $\rfidtag$ out-of-band. 
He then generates a random $x$, computes
$\encid = (u,v) = (\rfidtag \cdot \pubkey_{\domain}^x,g^x)$
and computes
$\kta = \enc{\kMA}{\rfidtag}$ using its own master access key $\kMA$, before
calling the method.
Note that this protocol is susceptible to hijacking and eavesdropping on the new owner's access
key, if the default session key $\bot$ is used (which is the case when the tag
has no owner). 

To transfer ownership of tag $\rfidtag$ from tag owner $\rfidtagowner$ to
domain $\rfidtagowner'$, a new entry for the new tag owner must be set in
$\accessset$ with a new encrypted tag identifier and a new diversified access
key (and in fact all other entries in the access set need to be deleted). 
The tag identifier does not change. This process is in fact a three party
protocol that is implemented by two method calls. The first runs as follows.
\begin{description}
\item[\func{TransferTagOwnership}($\caller$)]:\\
	verify $\owner = \caller$ ; \\
	$\accessset \assign \emptyset$ (hence $\owner = \bot$) ; 
\end{description}
Note that this function can only be executed in sessions of the authentic(ated) tag owner. 
After execution of this function, the session is \emph{not} terminated 
(i.e. the session key is \emph{not} reset). While in this state, the tag is 
shipped to the new owner $\rfidtagowner'$ and the values of the tag identifier $id$, 
the session key $s$ and the message counter $n$ are sent to $\rfidtagowner'$ out of band.
Then, $\rfidtagowner'$ calls $\func{TakeTagOwnership}$
(without prior authenticating and hence starting a new session!), 
thus becoming the new tag owner
(preferably when the old owner is out of reach so it cannot eavesdrop 
on the new values sent to the tag).

We note that the above described method might pose problems for domains that
need to take ownership for many tags, as e.g. electronics manufacturers or
retailers may do (see Scenario 1). They would face a practical problem of how
to determine which tag would be associated to which tag identifier and which
session key to use, which could
easily become an adminstrative nightmare. Also, it would be more in line with
Anderson's Duckling protocol~\cite{StaA99,stajano2000duckling-what-next} if
anyone can just take ownership of an unowned tag without any other
knowledge. For unowned (and unowned only) tags one could enable a method
that returns the unencrypted tag identifier. To transfer the ownership of many
tags using a single session key, one could extend the method
\func{transferTagOwnership} with an additional parameter $s$ to set the
session key on the tag to a fixed value.

\subsection{Granting access to a domain}
\label{ssec-grantingaccess}

To grant a domain $\domain$ access to a tag $\rfidtag$, its access set entry
$\accessset_\rfidtag[\domain]$ needs to be set with 
a new encrypted tag identifier and a new diversified access key.
This process is again a three party protocol that is implemented by two method calls. 
None of these methods require additional permission tokens to be executed.
The first method called (by the tag owner) is
\begin{description}
\item[\func{GrantTagAccess}($\caller,\domain$)]:\\
	verify $\owner = \caller$ ; \\
	$\accessset[\domain] \assign \bot$ ; 
\end{description}
The tag identifier, the value of the session key as well as the value of the message counter $n$
are sent to the domain $\domain$ out of band.
He subsequently calls (\emph{not} authenticating and starting a new session!)
\begin{description}
\item[\func{AcceptTagAccess}($\caller,\domain,\encid,\kta$)]:\\
	verify $\accessset[\domain] = \bot$ ; \\
	$\accessset[\domain] \assign (\encid,\kta, \id{false})$  ;
\end{description}
computing
$\encid = (u,v) = (\rfidtag \cdot \pubkey_{\domain}^x,g^x)$
and 
$\kta = \enc{\kMA}{\rfidtag}$ 
as in the case of \func{TakeTagOwnership}.
Note that the remarks made for \func{TakeTagOwnership} with respect to 
the need to communicate the tag identifier, apply here equally well.
Also, an improvement to these functions can be made if it would not be
necessary to have a pending session in between the calling of \func{GrantTagAccess} 
and \func{AcceptTagAccess} as a refusal to execute \func{AcceptTagAccess} 
would constitute a denial of service.
\begin{description}
\item[\func{RevokeTagAccess}($\caller,\domain$)]:\\
	verify $\owner = \domain$ ; \\
	$\accessset[\domain] \assign \bot$ ; 
\end{description}

\subsection{Re-encrypt identifiers}
\label{ssec-reencrypt-method}

The following two functions allow a reader to re-encrypt all encrypted tag
identifiers stored in $\accessset$. First the reader retrieves the current
encrypted tag identifiers in an array through the following method.
\begin{description}
\item[\func{ReEncryptGetIds}($\caller$)]:\\
	verify $\owner = \caller$ ; \\
	return a list of all encrypted tag identifiers in $\accessset$ ; 
\end{description}
The reader then computes the re-encryption of each of the entries in
$\accessset$ as described in Sect.~\ref{ssec-reencrypt}, creating a new array
$R$. Finally, to upload the new entries to the tag, it calls the following
method. 
\begin{description}
\item[\func{ReEncryptPutIds}($\caller,R$)]: \\
	verify $\owner = \caller$ ; \\
	store each entry in $R$ in the corresponding location in $\accessset$ ;
\end{description}
Both methods can only be called by the tag owner.
Alternatively, one could require that the caller owns a permission token to
call the method. 

\subsection{Managing objects}
\label{ssec-manage}

Managing an object involves the creation, deletion or update of the object on a
particular tag $\rfidtag$. These are handled by the following methods.

To install an object, one needs to call the following method on the object
manager object $\objmngrobj$. Depending on requirements, one may decide that
further permission tokens are necessary, or instead require a specific
permission token from the tag management object .
\begin{description}
\item[\func{InstallObject}($\caller,i,o,\symkey$)]:\\
	verify $\owner = \caller$ ; \\
	verify that object with name $i$ does not exist on the tag yet ; \\
	create a new object $o$ with name $i$ with class key $\symkey$ ;  
\end{description}
To update or delete an object, one needs to call one of the following methods
\emph{on the object to be updated or deleted}. Additional permission tokens 
from that
object may be required.
Only the owner of a tag can delete an object.
\begin{description}
\item[\func{UpdateObject}($\caller,i,o$)]:\\
	update object with name $i$ to $o$ ; 
\item[\func{UpdateClassKey}($\caller,i,\symkey$)]:\\
	update the class key of object with name $i$ to $\symkey$ ;  
\item[\func{DeleteObject}($\caller,i$)]:\\
	verify $\owner = \caller$ ; \\
	verify $i \neq \objmngrobj$ ; \\
	delete the object with name $i$ ; 
\end{description}

Note that by implementing object management this way, objects can only be
managed by domains that 
\begin{itemize}
\fixlistspacing
\item have access to the tag because they are a member of its access set
$\accessset$, and
\item have the correct permission token for the tag management object $\objmngrobj$,
issued using its class key $\objmngrkey$.
\end{itemize}
Note that the tag management object itself can also be updated this way (and in
particular its key), but cannot be removed or created.
When tags are created, a default tag management object is present on the tag.

Also note that neither the tag owner nor the owner of the tag management object
is capable of removing objects that they do not own, or do not have a delete permission for.
In order to prevent tags becoming unusable because of the multitude of objects installed on it,
one might consider to extend the functionality of \func{RelinquishTagOwnership} to include
the deletion of every object (except, of course, the tag management object) on the tag.

\section{Security analysis}
\label{sec-secanalysis}

We first give a security analysis of the authentication protocol from
Sect.~\ref{sec-auth} against the most important
security properties mentioned in that
section. We then analyse the security of the method invocation protocol
from Sect.~\ref{ssec-callingamethod}.

The adversary we consider has full control over the communication medium: he
can block, intercept, duplicate, modify and fabricate arbitrary messages. He
can, however, not create valid MACs for messages if he does not know the key,
and cannot encrypt or decrypt messages for which he does not know the symmetric
key. The adversary can corrupt arbitrary tags and hence can know their full state
including any keys they store. The adversary can also corrupt arbitrary
readers. However, such readers are known to be corrupted and the system is
notified of any such corruption.

Let $\secparam$ be the security parameter (implicitly defined by the size of
$G$ (see~\ref{ssec-reencrypt}) and the choice of the size of the symmetric keys).


We first prove the security of the authentication protocol.
\begin{lemma}
\label{lem-auth}
Let a reader from domain $\domain$ call the function
 $\id{AuthenticateR}(\kD,\KD,\kMA)$
which returns $(\sigma,\rfidtag')$. Let tag $\rfidtag$ call
$\id{AuthenticateT}()$ which returns $(\sigma',\domain')$. Then 
$\sigma=\sigma'$ only
if $\rfidtag = \rfidtag'$ and $\domain = \domain'$. No other entity not in
domain $\domain$ knows $\sigma$.
\end{lemma}
\begin{proof}
Consider the protocol in Fig.~\ref{fig-protenc-auth-2}.
Suppose $\sigma = (s \xor \bar{s}') = (s' \xor \bar{s}) =\sigma'$. 
Then the reader accepted the message
$\enc{\kta'}{q' \concat \bar{s}}$. Hence $\kta = \enc{\kMA}{\rfidtag'}$ as
computed by the reader equals $\kta'$. As
$\kta'$ is retrieved from $\accessset[\domain']$ and $\kMA$ is only known to
$\domain$ this proves $\domain = \domain'$

Similarly the tag must have accepted the message
$\enc{\kta}{\encid'\concat \curepoch \concat r' \concat q \concat \datetime \concat s}$
using its own key $\kta'$. Again
for $\kta = \enc{\kMA}{\rfidtag'}$
we must have $\kta'=\kta$. Because only $\rfidtag$ holds 
$\kta = \enc{\kMA}{\rfidtag}$ we must have $\rfidtag = \rfidtag'$.

To know $\sigma$ one needs to know both $s$ and $\bar{s}$. This requires one
to know $\kta$. Clearly $\rfidtag$ knows this. Otherwise, it requires one to
know $\kMA$ (and $\rfidtag$). This is only known to members of $\domain$. This
proofs the final statement of the lemma.
\qed
\end{proof}


Privacy after authentication or full re-encryption is guaranteed by the
following lemma.
\begin{lemma}
\label{lem-privacy}
Let $\rfidtag$ be a tag, whose tag identifier $\rfidtag$ for domain $\domain$
gets re-encrypted from $\encid$ to $\encid'$ (either by authentication or by a
full re-encryption). Let $\encid''$ be the encrypted tag identifier for domain
$\domain$ of an arbitrary tag $\rfidtag' \neq \rfidtag$. Then there exists no
adversary (that has no access to the private keys of domain $\domain$) with
resources polynomially bounded in $\secparam$ that can decide whether $\encid'$
and $\encid''$ or $\encid'$ and $\encid$ are encrypted tag identifiers of the
same tag.
\end{lemma}

\begin{proof}
In~\cite{golle2004reencryption} it is shown that, given our use of
ElGamal over our choice of group $G$, there does not exist an adversary with
resources polynomially bounded in $\secparam$ that can properly match the
re-encryptions of two ciphertexts with the original input ciphertexts. 
The adversary linking either $\encid$ or $\encid''$ with $\encid'$ would
trivially solve this problem too, and hence cannot exist either.
\qed
\end{proof}


Resilience to reader compromise is shown by the following lemma.
\begin{lemma}
A reader from domain $\domain$ reported stolen in epoch $\epoch$ cannot decide
whether two tags that have successfully authenticated with an unstolen reader
from domain $\domain$ in epoch $\epoch'> \epoch$ corresponds with a tag
observed before that authentication.
\end{lemma}
\begin{proof}
At the start of epoch $\epoch'$, we have $\curepoch = \epoch'$, and
all readers in domain $\domain$ that are not reported stolen
receive new epoch keys $(\kD',\KD')$ that are stored in $\eka[\curepoch]$. 
If a tag authenticates with this reader, according to the protocol, it receives
a new encrypted identifier encrypted with the keys
$(\kD',\KD')$. Let two tags meet such a reader, obtaining encrypted 
tag identifiers $\encid'_a$ and $\encid'_b$ in exchange for their old
identifiers $\encid_a$ and $\encid_b$.
If subsequently these tags meet a reader from domain $\domain$ that was
reported stolen in epoch $\epoch < \epoch'$, this reader does not own the
key pair $(\kD',\KD')$ and hence cannot decrypt $\encid'_a$ or $\encid'_b$. 
Therefore, by Lemma~\ref{lem-privacy}, the reader cannot link the
previous encrypted identifiers $\encid_a$ and $\encid_b$.
\qed
\end{proof}


Finally, we show security of the method invocation protocol.
\begin{lemma}
A tag  $\rfidtag$ only executes a method $\method$ of class $\class$ 
with class key $\kc$ if a reader in domain $\domain$ 
with
\begin{itemize}
\fixlistspacing
\item $\accessset_\rfidtag[\domain] \neq \bot$ when it starts the session, and
\item permission token $\pcfod = \enc{\kc}{\method,\domain,\maxtime}$ 
with $\maxtime > \now_\rfidtag$ (when the permission is verified)
\end{itemize}
issued the command to the execute this method in the session it started. Moreover,
the tag will do so at most once.
\end{lemma}
\begin{proof}
Checking the protocol, we see that a tag $\rfidtag$ executes method $\method$
on class $\class$ with class key $\kc$ when
\begin{itemize}
\fixlistspacing
\item it receives a message correctly encrypted and
mac-ed with its session key $s$, containing the parameters and the expected
message counter $m+1$, and before that
\item has received a message correctly encrypted and
mac-ed with its session key $s$, containing 
$\method$, $\class$, $\maxtime$ and 
a permission token $\pcfod = \enc{\kc}{\method,\domain,\maxtime}$ with
$\maxtime > \now_\rfidtag$, and the expected message counter $m$.
\end{itemize}
The authentication protocol guarantees (see Lemma~\ref{lem-auth}) that
only if $\domain$ is a member of $\accessset_\rfidtag$
when starting a session, the reader and the tag 
share the same session key $s$.
Therefore, in the current session the tag only accepts messages
constructed by such a  reader in domain $\domain$. This proves that
it must have issued the command to the execute this method in the session
it started, and also that it held the appropriate permission token.
Moreover, due to the use of message counters, the current session only accepts
a particular message encrypted for this session at most once.
This proofs the final statement of the lemma.
\qed
\end{proof}

\section{Concluding remarks and further research}
\label{sec-concl}

We have presented a model for a fine grained and dynamic management of access
permissions to RFID tags, and we have presented privacy friendly protocols
efficiently implementing this model. This efficiency is achieved by avoiding a costly key search algorithm at the reader side. The price to pay is a little less privacy: tags can be traced in between successful authentications by legitimate readers. However, this is mitigated quite effectively by giving tag owners the possibility to re-encrypt tag identifiers at any point in time.

Although the model accommodates a multitude
of use cases, in the course of this research we have identified several
capabilities that our current implementation lacks. 
\begin{itemize}
\fixlistspacing
\item Access to tags and objects is bound to specific domains.
  A domain with certain permissions cannot delegate them to another domain.
  Instead new permissions have to be requested from the tag owner and the class
  owner. 

\item Although access to a \emph{tag} can be revoked instantaneously, permission
  tokens to access \emph{objects} cannot be revoked (although their validity
  can be constrained by using short validity periods).

\item Another approach to limit validity of permissions is to issue one-time
  only permission tokens that can be used exactly once to call a particular
  method on an object.

\item Domains are granted access to specific tags one by one by the respective
  tag owners. Permission tokens to call a method on an object are however not
  tag specific (unless each object of the same class is given a separate class
  (or rather object) key.

\item The distinction between a permission to access a tag and a permission to
  call a method on an object is confusing and perhaps unfortunate. 
  This distinction arises from two factors. First, access to a tag is issued by
  the current owner, and is maintained on the tag to allow immediate
  revocation of access. Moreover, the privacy friendly authentication protocol
  needs to know which domains have access to the tag -- hence tag related
  access control decisions are taken at a lower layer than object related
  access control decisions.
\item Finally, to re-encrypt an identifier, one needs to own the corresponding access key. This severely limits the options for owners to re-encrypt their tags. On the other hand, not requiring such an access key puts tags wide open
to denial-of-service attacks that feed them with bogus identifiers.
\end{itemize}
Further research is
necessary to see whether these capabilities are truly necessary in real-life
applications, and, if so, how these capabilities can be added efficiently.
We welcome discussion and feedback on these issues.

\bibliography{sensorpets}

\end{document}